# The Impact of Cooling Rate on the Tensile and Cyclic Stress-Strain Characteristics of Different Solder Alloys at Nanoscale

Sadib Fardin, Md. Jawarul Moresalein, Toushiqul Islam, Abrar Faiyad and Mohammad Motalab*

*Department of Mechanical Engineering, Bangladesh University of Engineering and Technology, Dhaka - 1000, Bangladesh*

sadib.fardin@gmail.com, moresaleinayon1998@gmail.com, xsizan.03@gmail.com, afaiyad@ucmerced.com, abdulmotalab@me.buet.ac.bd (*corresponding author)

**ABSTRACT**

In recent years, lead-free solder alloys based on tin, silver, or copper have gained popularity over lead-based solder alloys due to their improved mechanical and electrical properties and their non-toxic nature. In our previous studies, we examined the stress-strain behavior of SAC305 under varying cooling rates. This study extends our investigation to various lead-free solder materials, including Sn, Sn-Ag, and SAC305, to compare their relative mechanical and cyclic properties. The cooling rate plays a crucial role in microstructure and mechanical properties. Therefore, we employed molecular dynamics to model the atomistic behavior. Initially, the models were melted at a constant rate and then cooled at various rates, including 2.5 K/ps, 10 K/ps, 50 K/ps, and 100 K/ps. Additionally, exponential cooling was used to replicate real-world cooling scenarios such as air cooling, water cooling, and furnace cooling. We utilized a set of modified embedded atomic model (MEAM) interatomic potentials for the tensile test and cyclic loading. The tensile test was conducted until fracture occurs at a strain rate of $1 \times 10^9$ s$^{-1}$. Furthermore, we investigated the cyclic loading behavior within a strain range of -10% to 10% for 10 cycles. The results indicated that cooling rates significantly influenced mechanical properties, with slower rates (2.5 K/ps and 10 K/ps) showing substantial differences, while the differences between higher rates (50 K/ps and 100 K/ps) were less pronounced. The ultimate strength, Young's modulus, modulus of resilience, and coefficient of thermal expansion exhibited a negative correlation with increasing cooling rates, while the modulus of toughness increased, indicating improved impact resistance. To assess energy dissipation during cyclic loading, we examined the hysteresis loop area and stress amplitude. After a certain number of cycles, the energy lost during each cycle reached a stable level.

**NOMENCLATURE**

| | |
|---|---|
| $E$ | Total energy |
| $F$ | Embedding energy |
| $m$ | Mass |
| $P$ | Pressure |
| $r$ | Interatomic distance |
| $T$ | Temperature |
| $v$ | Velocity |

*Greek symbols*

| | |
|---|---|
| $\rho$ | Atomic electron density |
| $\phi$ | Pair potential interaction |
| $\sigma$ | Stress |
| $\tau$ | Time constant |
| $\epsilon$ | Strain |
| $\Omega$ | Atomic volume |

*Superscript*

| | |
|---|---|
| $\alpha, \beta$ | Atomic indices |

*Subscript*

| | |
|---|---|
| $i.j$ | Indices in Cartesian coordinate system |

*Abbreviations*

| | |
|---|---|
| CTE | Coefficient of thermal expansion |
| UTS | Ultimate tensile strength |
| YM | Young's modulus |

## 1. INTRODUCTION

The electronics industry is shifting towards lead-free soldering techniques due to environmental and health concerns [1]. To replace lead, various lead-free materials, including Sn, Sn-Ag, and Sn-Ag-Cu (SAC), have been developed and their mechanical and thermal properties extensively examined to ensure their reliability [2-4]. Lead-free solder materials exhibit exceptional qualities, such as lower melting points [5], excellent wetting and solderability [6], improved thermal and mechanical properties [7, 8], compatibility with common electronic components [9, 10], and relatively high electrical conductivity [11]. Consequently, they play a crucial role in applications such as solder bumping for flip chip connections, die bonding, mounting microelectronic devices, surface mount technology (SMT), and plated-through-hole (PTH) assembly techniques [12-14]. However, persistent issues remain, including copper dissolution, excessive intermetallic compounds, relatively high cost, and tin whiskers [15-17]. Sn-Pb solder remains popular due to its lower melting point, reduced surface tension, and its ability to inhibit the transformation of white tin to gray by acting as a solvent element [18].

The tensile properties of lead-based and lead-free solder are compared under different temperatures and strain rates [19]. The issue of fatigue loading is of significant importance in Micro-electromechanical systems (MEMS) applications throughout their operational lifespan [20]. Pang et al. [21] investigated the impact of temperature and strain rate on the fatigue life of solder joints using finite element methods. Lall et al. [22] conducted a study on the characterization of SAC solder after extended storage at low temperatures and high strain rates to determine nine anand parameters from tensile data. Chowdhury et al. [23] employed a water-quenched solidification profile to analyze the mechanical properties of doped SAC solder materials. Basaran et al. [24] explored the material response to concurrent vibration and thermal loading. Wiese et al. [25] used cyclic triangular strain waves to assess the durability of solder materials in various applications. The fatigue life of SAC and Sn-Pb solder has been compared for different package types by Schubert et al. [26].

Numerous numerical studies of lead-free solder materials have been conducted in recent years. SAC alloys with varying silver (Ag) content were computationally evaluated at different temperatures to assess their thermomechanical properties [27]. Zhang et al. [28] studied the growth of tin whiskers and intermetallic compound evolution in SAC307 using computational methods. Zhang et al. [29] employed numerical simulations to analyze atom diffusivity at the Ag3Sn-βSn interface in Sn-Ag solder, aiming to understand vital micro-mechanisms governing void formation and improved solder joint properties. Li et al. [30] conducted numerical and experimental investigations into the shear deformation of Sn-based solder joints. Chen et al. [31] demonstrated cyclic stress-strain behavior in Ni-based single crystal alloys by modulating temperature and strain range under cyclic loading. Alvi et al. [32] conducted cyclic and tensile tests to study dislocations and defects in Gold-Silver core-shell systems. Nguyen et al. [33] examined the effects of temperature and strain rate on cyclic plasticity in AlCrCuFeNi HEA, revealing interactions between partial dislocations and induced lattice disorders during deformation.

The solidification process can be analyzed using various computational techniques, including Monte Carlo simulation, finite element simulation, and phase-field simulation [34-36]. The microstructure, a key determinant of mechanical properties, is directly influenced by the solidification process [37-39]. Molecular dynamics can be employed to observe the microstructure at each stage of quenching and comprehend the solidification process [40]. Recently, Li et al. [41] investigated the nucleation process and grain growth of AlCoCrCuFeNi across a wide range of cooling rates. Shen et al. [42] rapidly quenched aluminum to examine glass formation and microstructure. In another study by Li et al. [43], different structures resulting from varying cooling rates were correlated with tungsten strength. Shu et al. [44] explored substructure formation in AlN/FeNiCrCo composites during rapid solidification via selective laser melting, revealing subgrain boundaries and a dense dislocation network originating from heat transfer and lattice mismatches.

To the best of our knowledge, no numerical analysis has been conducted at the nanoscale level to examine the impact of quenching rates on different solder alloys. In this study, we melted Sn, Sn-Ag, and SAC305 above their melting points at a constant cooling rate and then quenched them at various rates, including 2.5 K/ps, 5 K/ps, 50 K/ps, 100 K/ps, and an exponential cooling rate. The resulting models, after the solidification process, were subjected to uniaxial tensile and fatigue loading. We assessed changes in mechanical and thermal properties resulting from varying cooling rates in solder alloys, evaluating parameters such as ultimate tensile strength (UTS), Young's modulus (YM), modulus of toughness, and modulus of resilience. Additionally, the models of solder alloys after solidification underwent cyclic loading analysis to assess the hysteresis loop area per cycle and stress amplitude. This study employed the modified embedded atomic model (MEAM) potential [45, 46].

## 2. METHODOLOGY

### 2.1 MEAM POTENTIAL

The current study utilizes the Large-scale Atomic/Molecular Massively Parallel Simulator (LAMMPS) for performing molecular dynamics simulations, employing a modified embedded atom method (MEAM) to describe interatomic bonding. MEAM has proven to provide higher accuracy in characterizing intermetallic interactions when compared to the EAM potential, mainly due to its inclusion of angular forces. The force field employed in this study comprises two primary energy components: the embedding energy potential and the pair potential. Following the MEAM formulation, the total energy of all atoms within a system, denoted as E, can be expressed as:

$$E = \sum_i \left\{ F_i(\bar{\rho}_i) + \frac{1}{2} \sum_{i \neq j} \phi_{ij}(r_{ij}) \right\}, \tag{1}$$

Here, the parameter $F$ represents the embedding energy, which is dependent on the electron density of the atom denoted by $\rho$ and pair potential interaction is denoted by $\phi$. The summation of the pair potential interaction is conducted over all neighboring atoms $j$ of atom $i$ that are located within a specified cutoff distance.

The relevant literature has previously established and documented the MEAM parameters for binary materials consisting of Sn, Ag, and Cu. Table 1 presents the established parameters corresponding to the solder materials for the current investigation.

**Table 1**: MEAM parameters of Sn, Sn-Ag, SAC305 [47-49]

| Parameters | Sn | Ag | Cu | Sn-Ag | Ag-Cu | Cu-Sn |
|---|---|---|---|---|---|---|
| $E_c$ (eV) | 3.08 | 2.85 | 3.54 | 3.14 | 3.5 | 2.8 |
| $r_0$ (Å) | 3.44 | 2.92 | 2.54 | 2.58 | 2.21 | 2.36 |
| $\alpha$ | 5.09 | 5.89 | 5.1 | 5.8 | 5.78 | 5 |
| A | 1.12 | 1.06 | 1.07 | | | |
| $\beta(0)$ | 5.42 | 4.45 | 3.63 | | | |
| $\beta(1)$ | 8 | 2.20 | 2.20 | | | |
| $\beta(2)$ | 5 | 6 | 6 | | | |
| $\beta(3)$ | 6 | 2.20 | 2.20 | | | |
| $t(0)$ | 1 | 1 | 1 | | | |
| $t(1)$ | 3 | 5.541 | 3.138 | | | |
| $t(2)$ | 5.707 | 2.45 | 2.494 | | | |
| $t(3)$ | 0.3 | 1.288 | 2.95 | | | |

## 2.2 ATOMIC STRUCTURE MODELING

The present investigation entails the modeling of four distinct nanostructures composed of different solder alloys, specifically Sn, Sn-Ag, and SAC305. The atomic model for β-Sn is constructed using Atomsk software within a box with dimensions of 288.7 x 72.17 x 72.55 (Å), as depicted in Figure 1, with atoms color-coded based on their respective atom type. Subsequently, the Sn atoms undergo a random replacement process with Ag and Cu, according to their respective proportions, as also illustrated in Figure 1. For fatigue tests, the dimension along the x-axis is scaled down to a 2:1 ratio, while the dimensions along the y and z axes remain unchanged. This aspect ratio reduction is implemented as a precaution against potential buckling issues that may arise during the compression process. The atomic structures utilized for tensile tests consist of 81,920 atoms, while those employed for fatigue tests comprise 40,960 atoms.

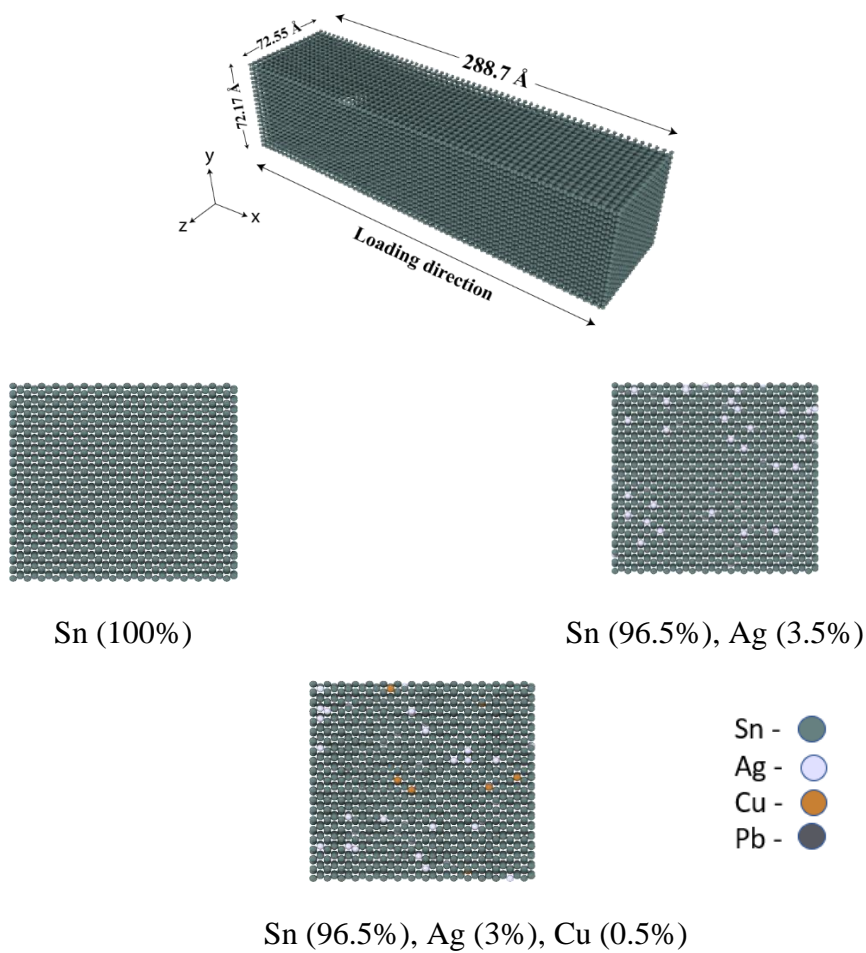

**Figure 1**: The atomic model of the β-Sn and illustrations of the three different materials with their proportions.

## 2.3 MD SIMULATIONS

Molecular dynamics simulations of quenched solder alloys were carried out using the open-source software LAMMPS [50], and the resulting data were visualized with OVITO [51]. To maintain computational efficiency and accuracy, a fixed time step of 1 fs was employed. Energy equilibration was conducted for 12.5 picoseconds using an NVE ensemble with a Langevin thermostat. The simulations were performed in the NPT ensemble, utilizing the Langevin thermostat method to maintain a constant temperature (T) and the Berendsen barostat method to ensure constant pressure (P). Subsequently, the systems were relaxed in the NPT ensemble for 12.5 ps. The temporal variation of temperature during the melting and solidification processes is illustrated in Figure 2.

To melt the materials, all systems were heated from 300 K to 1200 K at a constant heat rate of 10 K/ps (A-B). The system was held at 1200 K for 12.5 ps (B-C) to achieve complete liquefaction. Subsequently, the system was cooled down to 300 K at various cooling rates: 2.5 K/ps (C-D), 10 K/ps (C-E), 50 K/ps (C-F), and 100 K/ps (C-G). An exponential cooling profile (C-H) was also employed to mimic the cooling profiles observed in air cooling, water bath cooling, or furnace cooling [52]. In this case, the temperature follows Equation 2 and is maintained by a loop of the NPT ensemble. After quenching, the systems were equilibrated at 300 K.

A strain rate of $1 \times 10^9$ s$^{-1}$ was applied to all samples along the x-direction within the NVT ensemble until fracture occurred. For cyclic loading, all systems were subjected to -10% to 10% strain of their initial length to induce plastic deformation in the solder alloys. Cyclic loading was applied for 10 cycles to analyze the area under the stress-strain curves. Stresses in the structures were calculated using the virial stress theorem [53], following Equation 3.

$$T = 1200e^{-\tau n} \qquad (2)$$

$$\sigma_{ij}^{\alpha} = \frac{1}{\Omega^{\alpha}} \left( \frac{1}{2} m^{\alpha} v_i^{\alpha} v_j^{\alpha} + \sum_{\beta=1,n} r_{\alpha\beta}^{j} f_{\alpha\beta}^{i} \right) \qquad (3)$$

In Equation 2, $\tau = 0.03$ is the time constant of the temperature profile and $n$ is the number of step in the loop. In Equation 3, $i$ and $j$ are the indices in the general Cartesian coordinate system, $\alpha$ and $\beta$ are the atomic indices, $m^{\alpha}$ and $v^{\alpha}$ denote the mass and velocity of the atom, $r_{\alpha\beta}^{j}$ is the distance between the atoms $\alpha$ and $\beta$, $\Omega^{\alpha}$ is the atomic volume of atom α.

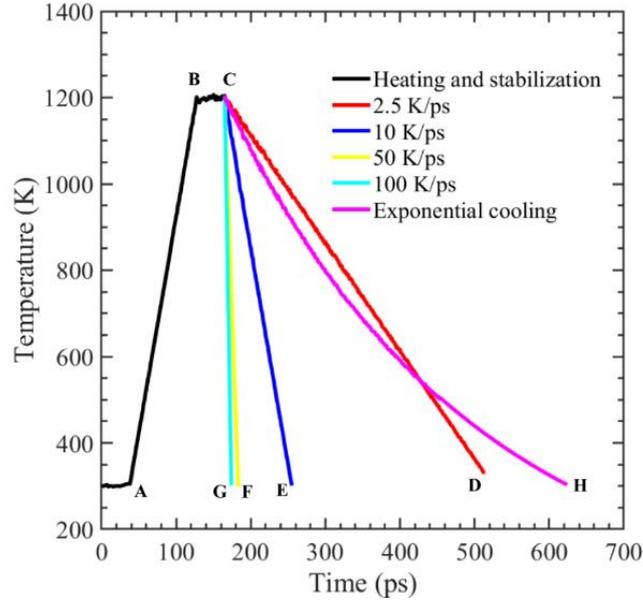

**Figure 2**: Temperature profile during heating and cooling of the structure.

## 3. METHOD VALIDATION

During the heating process, the system's temperature is elevated from 300 K to 1200 K, surpassing the melting point of SAC305. To validate the precision of our model, we plotted the total energy of the system against the temperature change, as depicted in Figure 3, to pinpoint the melting region. Initially, the model was in the solid phase within region (i). Subsequently, a significant increase in the slope of the total energy curve was observed between points A and B in region (ii), indicating the commencement of bond breaking due to a structural phase transition. Moreover, a distinct change in slope is noticeable, starting at point B across region (iii), signifying the complete liquefaction of the substance. The initial transition point in the slope was observed at approximately 490 K, which closely aligns with the experimental finding of 494 K [54], thus confirming the accuracy of our model.

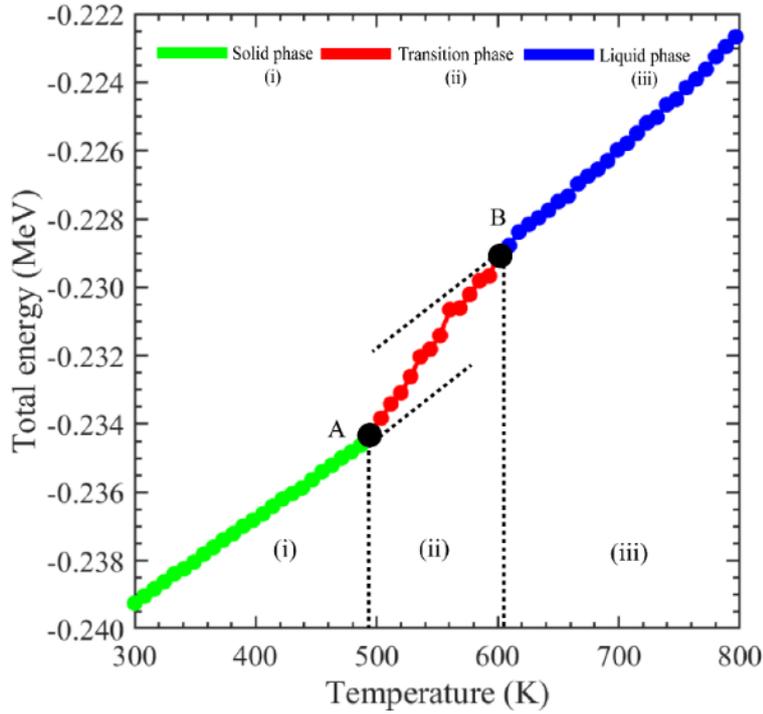

**Figure 3**: Change in the overall energy of the SAC305 configuration is observed as the temperature is raised.

Notable previous studies' results are listed in Table 2 for validating the computational method that is used in present work.

**Table 2**: Validation against the experimental studies by CTE

| Materials | Previous Experiments ($10^{-6}$ K$^{-1}$) | This Study ($10^{-6}$ K$^{-1}$) |
|---|---|---|
| Sn | 24-29.5 [55] | 23.8 |
| Sn-Ag | 19.4-23.2 [56] | 25.9 |
| SAC305 | 17.6-23.6 [56] | 22.1 |

## 4. RESULTS AND DISCUSSION

### 4.1 UNIAXIAL TENSILE LOAD

The stress-strain curves for Sn, Sn-Ag, and SAC305 at a constant strain rate of $1 \times 10^9$ s$^{-1}$ are plotted in Figure 4 for various cooling rates (2.5 K/ps, 10 K/ps, 50 K/ps, 100 K/ps, and exponential cooling). Noticeable differences exist among the stress-strain curves. For higher cooling rates, such as 50 K/ps and 100 K/ps, there is no significant alteration in the stress-strain behavior. However, as the cooling rate escalates from 2.5 K/ps to 50 K/ps, substantial changes occur in the mechanical properties. The stress-strain response indicates elastic deformation occurring around 5-6% strain, followed by continued plastic deformation.

The variations in Ultimate Tensile Strength (UTS) and Young's Modulus (YM) in the stress-strain curves for all materials at cooling rates of 2.5 K/ps, 10 K/ps, 50 K/ps, and 100 K/ps are depicted in Figure 5 and 6. It is observed that an increase in cooling rate leads to a reduction in the UTS of the lead-free solder alloys. When the cooling rate is raised from 2.5 K/ps to 100 K/ps, the UTS decreases from 6.42 GPa to 5.60 GPa, 6.83 GPa to 6.04 GPa, and 6.90 GPa to 10.91 GPa, indicating a reduction of 12.77%, 11.55%, and 10.91% for Sn, Sn-Ag, and SAC305, respectively.

Similarly, the YM experiences a decrease with an increasing cooling rate. For Sn, the YM decreases from 6.42 GPa to 5.60 GPa, reflecting a reduction of 12.77%. For Sn-Ag, the YM decreases from 6.83 GPa to 6.04 GPa, indicating an 11.55% reduction. In the case of SAC305, the YM decreases from 6.90 GPa to 10.91 GPa, signifying a 10.91% reduction. The increase in cooling rate limits the time available for atom arrangement, leading to the creation of defects, which, in turn, results in a decline in UTS and YM for all lead-free solder alloys.

Additionally, it's noteworthy that both the UTS and YM show little variation between the cases of 50 K/ps and 100 K/ps. This phenomenon can be explained by the transformation of the microstructure from crystalline to amorphous characteristics due to the rapid solidification process, and in such cases, microstructures do not change significantly.

Furthermore, among all materials, SAC305 exhibits the maximum UTS, while Sn displays the lowest UTS. Similarly, in terms of YM, the SAC305 alloy also boasts the highest YM, while pure Sn possesses the lowest YM. This can be attributed to the strong bonds formed between Ag, Cu, and Sn atoms in SAC305 solder, resulting in increased resistance to deformation and consequently higher UTS and YM for the SAC305 alloy [57].

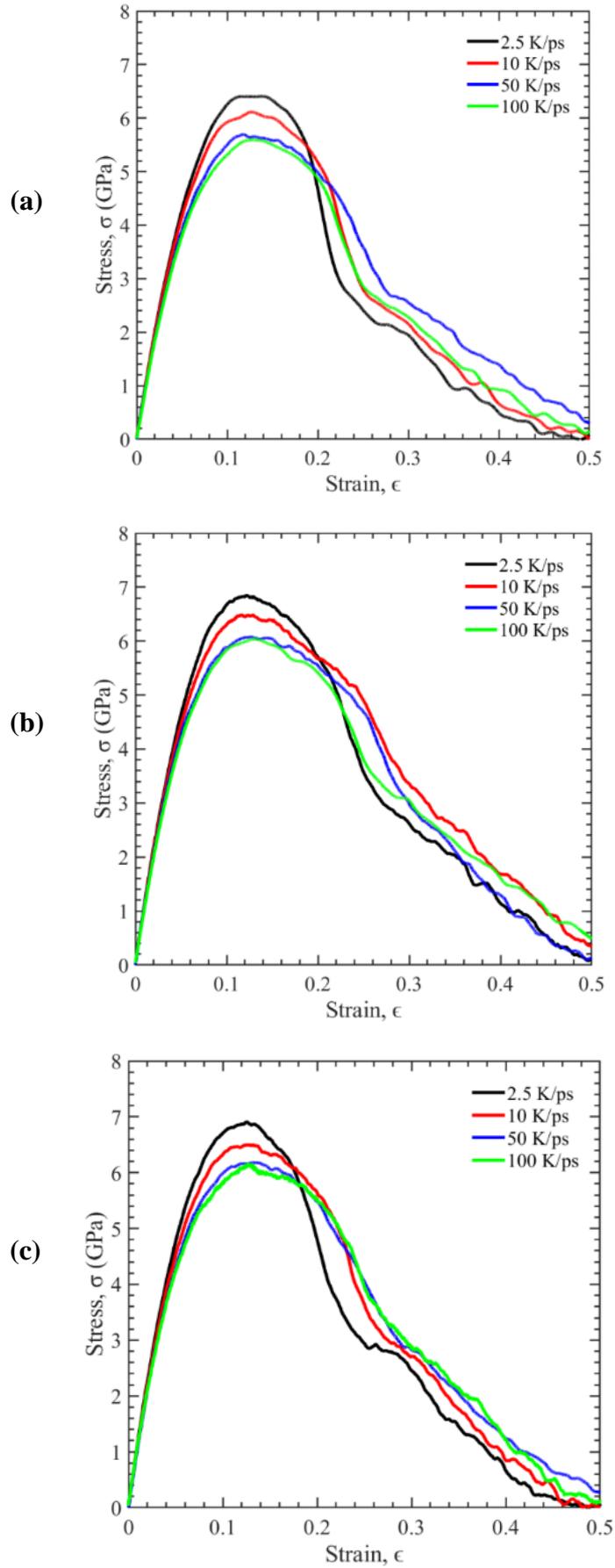

**Figure 4**: Stress-strain curves for different cooling rates of (a)Sn, (b) Sn-Ag, (c) SAC305

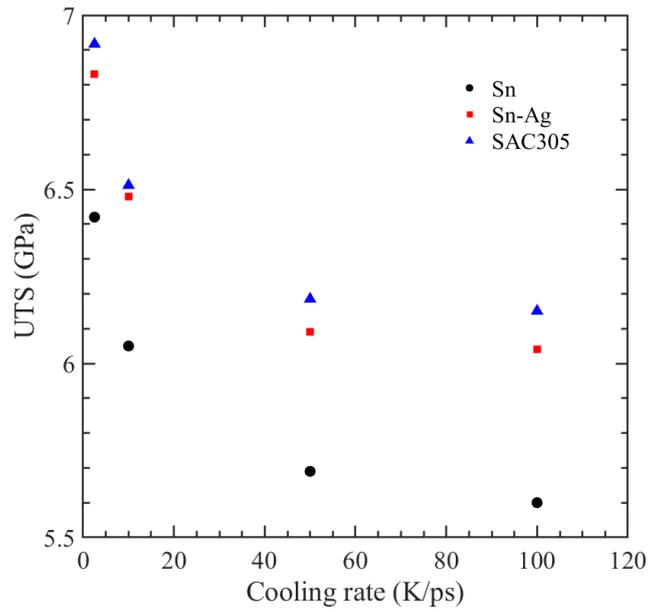

**Figure 5**: Variation of UTS with cooling rate

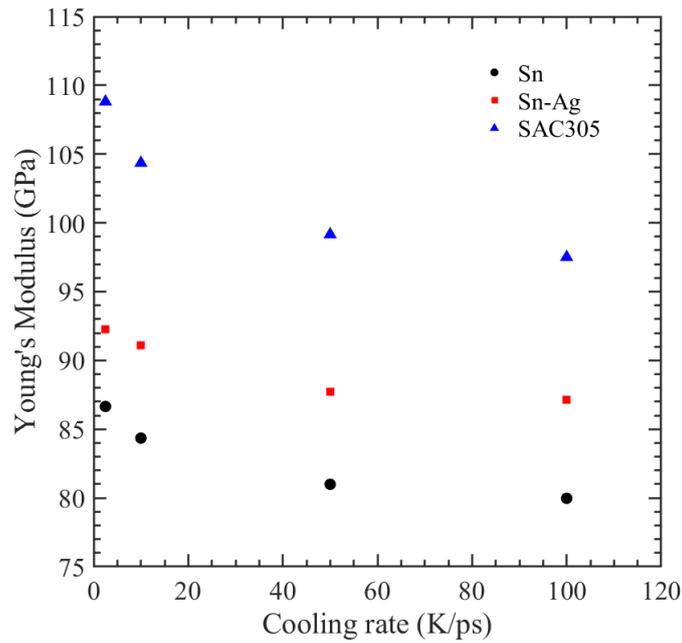

**Figure 6**: Variation of YM with cooling rate

The area under the curve is calculated to determine the modulus of toughness, and the results are presented in Figure 7 against the cooling rates. As the cooling rate increases, there is a corresponding increase in the modulus of toughness, indicating an improved capacity to absorb energy. It increases by 10.42%, 10.19%, and 6.96% for Sn, Sn-Ag, and SAC305, respectively, when the cooling rate is raised from 2.5 to 100 K/ps. As mentioned earlier, the higher density of defects such as dislocations and grain boundaries that occur with faster cooling rates can serve as sites for energy dissipation during deformation. Additionally, the formation of intermetallic compounds can contribute to increased toughness by providing additional sites for energy absorption. The addition of Cu to solder materials can potentially decrease their ductility by forming larger and more brittle intermetallic compounds. These intermetallic compounds can act as stress concentration points, making the solder alloy more prone to localized failure by reducing the strain energy absorption capacity of SAC305 [58, 59]. Thus, among all the materials, Sn-Ag has the highest modulus of toughness, indicating it is more ductile than the other materials, while Sn has the lowest.

The graph shown in Figure 8 illustrates the modulus of resilience of all alloys with increasing cooling rates. This parameter characterizes the material's ability to absorb and subsequently restore elastic strain energy during elastic deformation. In contrast to the modulus of toughness, the modulus of resilience decreases with increasing cooling rates. When the cooling rate is increased from 2.5 K/ps to 100 K/ps, there is a reduction of 10.42% for Sn, 10.19% for Sn-Ag, and 6.96% for SAC305. This suggests that the ability of the material to absorb and store elastic energy decreases with higher cooling rates. The incorporation of Ag and Cu into the composition enhances the strength of solder alloys, resulting in SAC305 having the highest modulus of resilience and pure Sn having the lowest modulus of resilience among all materials and cooling rates. Notably, both the modulus of toughness and modulus of resilience for 50 K/ps and 100 K/ps are nearly identical, further indicating the amorphous structure at higher cooling rates (50 K/ps and 100 K/ps).

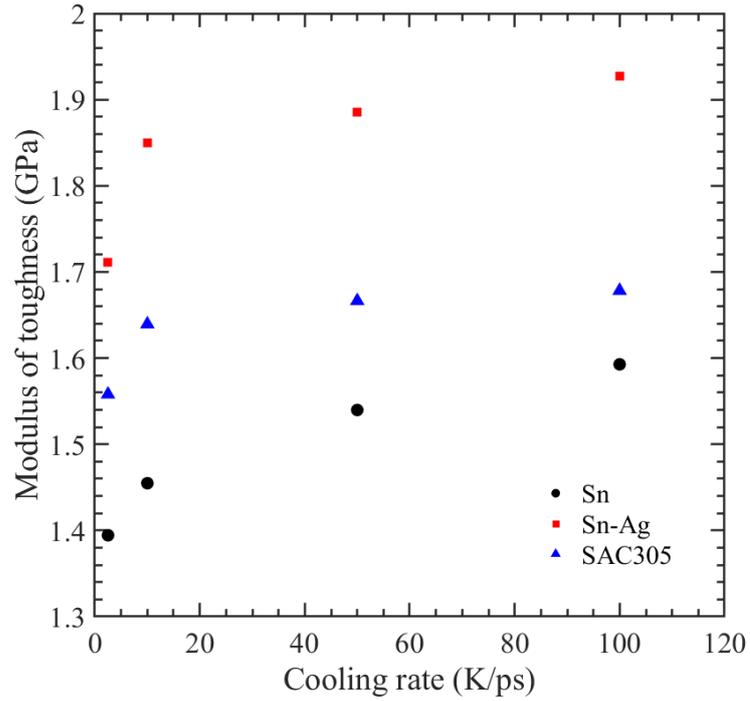

**Figure 7**: Variation of modulus of toughness with cooling rate

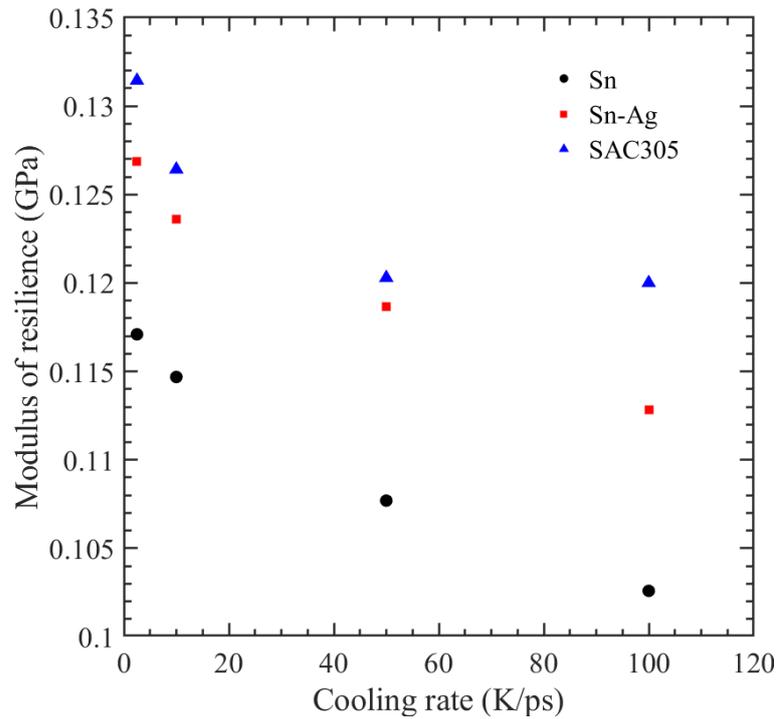

**Figure 8**: Variation of modulus of resilience with cooling rate

Figure 9 illustrates the relationship between the coefficient of thermal expansion (CTE) and cooling rates for all the materials. As the material's temperature decreases under NPT conditions, the simulation model size also decreases. The recorded size is used to calculate the CTE based on Equation (4). The size used in the equation is determined after the temperature reaches 200 ºC. For all the materials discussed, the CTE value decreases as the cooling rate increases. At higher cooling rates, the atoms have insufficient time to contract, which is a contributing factor to the decreasing trend in CTE. It's important to note that the CTE values at 50 K/ps and 100 K/ps are not negligible, unlike the properties discussed in the preceding section.

$$\alpha_v = \frac{1}{V}\frac{dV}{dT} \tag{4}$$

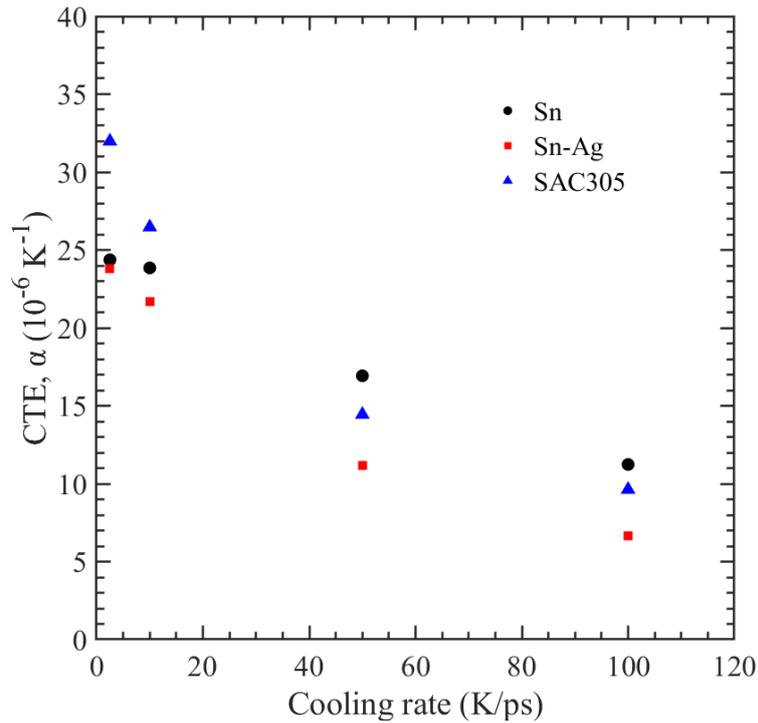

**Figure 9**: Variation of coefficient of thermal expansion (CTE) over different cooling rates

**4.2 EFFECT OF EXPONENTIAL COOLING**

When a material undergoes air cooling, water cooling, or furnace cooling, the temperature decay follows an exponential [60] rather than a linear pattern. To replicate a similar scenario, this study employs exponential cooling for all materials. The stress-strain behavior after exponential cooling to room temperature is illustrated in Figure 10, and the properties are compared to those obtained at a cooling rate of 2.5 K/ps. The results of exponential cooling closely resemble those of a cooling rate of 2.5 K/ps, with the former showing a slightly higher ultimate tensile strength. The UTS, YM, modulus of toughness, and modulus of resilience for all materials are presented in Table 3.

Table 3: Data from tensile test of exponential cooling

| Materials | UTS (GPa) | YM (GPa) | Modulus of toughness (GPa) | Modulus of resilience (GPa) |
|---|---|---|---|---|
| Sn | 6.46 | 83.47 | 1.50 | 0.104 |
| Sn-Ag | 6.51 | 91.94 | 1.69 | 0.118 |
| SAC305 | 6.91 | 108.64 | 1.62 | 0.120 |

(a)
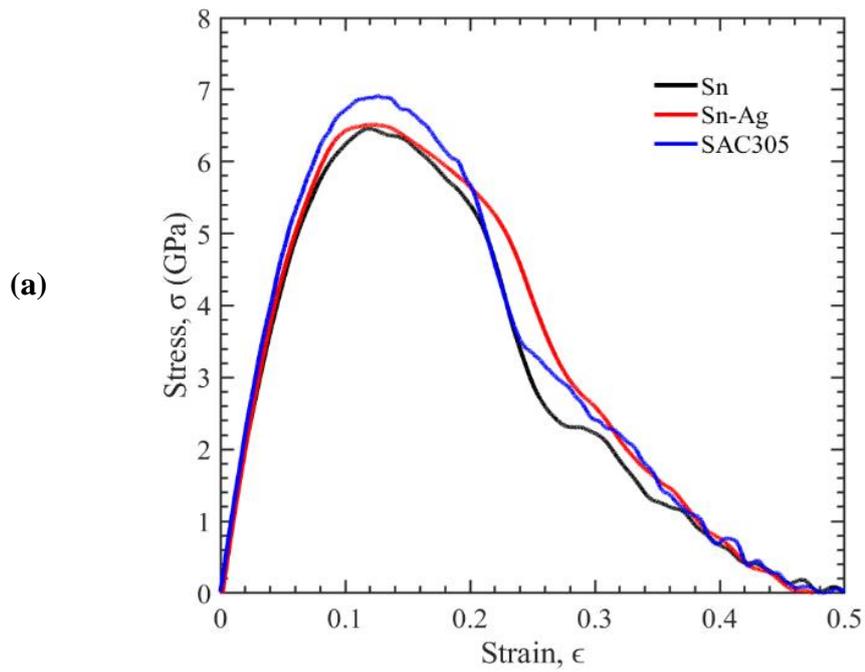

(b)
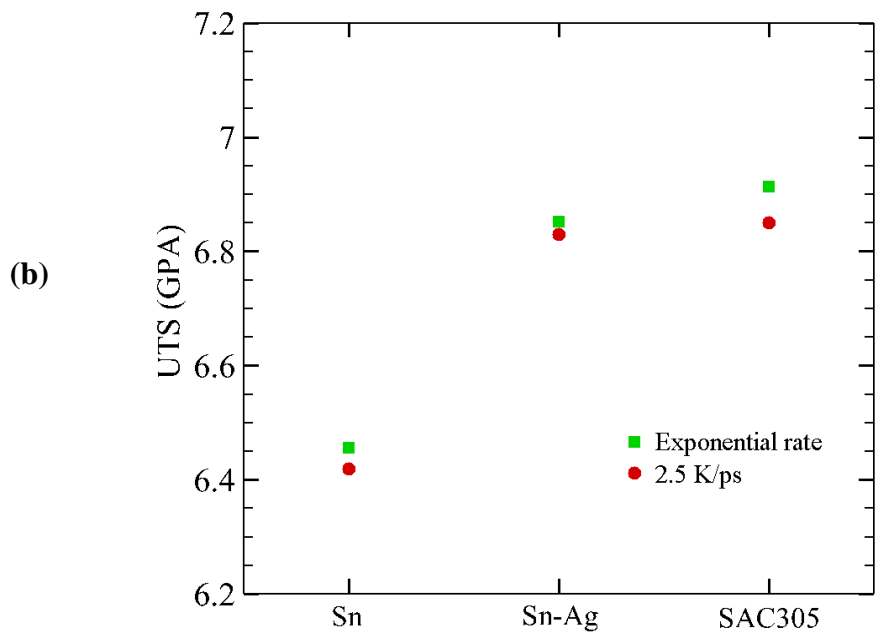

**Figure 10**: (a) Stress vs strain curve for different materials when cooled exponentially. (b) Differences in the UTS when cooled linearly (2.5 K/ps) and exponentially

## 4.3 EFFECTS OF CYCLIC LOADING

Figures 11, 12, and 13 illustrate the stress-strain behavior of Sn, Sn-Ag, and SAC305, respectively, under cyclic loading conditions for all cooling rates. Sn, Sn-Ag, and SAC305 are subjected to cycles of -10% to 10% of their original length at 2.5 K/ps, 10 K/ps, 50 K/ps, and 100 K/ps at 300 K. Notably, the results at 100 K/ps are comparable to those at 50 K/ps, mirroring the behavior observed in the tensile test. In lead-free solder alloys, elastic deformation occurs within the range of 6-8% strain, followed by plastic deformation up to 10% strain. Subsequently, the loading direction is reversed, marking the start of compression, but no elastic deformation is observed in the following cycles. This behavior aligns with experiments conducted by Mustafa et al. [60].

Figure 14 depicts the hysteresis loop area for the second, fifth, eighth, and tenth cycles at all cooling rates mentioned. The hysteresis loop area quantifies the energy dissipated per cycle and correlates directly with the material's damage accumulation during each loading-unloading cycle [61]. Variations in hysteresis loop areas are evident across all cooling rates and solder alloys. However, these differences progressively diminish in subsequent cycles, converging to a certain value. This indicates that saturation is reached in a shorter number of cycles, primarily due to stress redistribution within the materials. Another significant observation is that the area increases with each cycle, reflecting cumulative degradation in the material, such as fatigue or plastic deformation, which can cause the hysteresis loop area to expand. Figure 15 provides a clearer illustration of the area variation caused by cooling rates for the 2nd, 5th, and 10th cycles. Among all the materials, SAC305 exhibits the highest damage accumulation, while Sn has the lowest during the 2nd cycle. However, in the 5th and 10th cycles, the areas for Sn-Ag and SAC305 become similar.

The stress amplitude for lead-free solder materials has been calculated using Equation (5). Stress amplitude is commonly employed to predict the fatigue endurance of a material, with fatigue life being inversely proportional to stress amplitude. As the stress amplitude decreases, the material's fatigue life increases. Figure 16 illustrates the evolution of stress amplitude over time for different cooling rates. Stress amplitude diminishes as the number of cycles increases, primarily due to cyclic softening [62]. Cyclic softening occurs as plastic deformation and the accumulation of plastic strain become more prominent in subsequent cycles. The stress amplitude approaches a saturation point over the number of cycles, with lower cooling rates like 2.5 K/ps and 10 K/ps requiring more cycles to reach saturation. Higher cooling rates, such as 50 K/ps and 100 K/ps, reach the saturation point more quickly.

$$Stress\ Amplitude = \frac{\sigma_{max} - \sigma_{min}}{2} \quad (5)$$

To understand why the hysteresis loop converges to a certain value, the radial distribution function (RDF) for Sn, Sn-Ag, and SAC305 at different cooling rates is displayed in Figure 17 for the 2nd and 10th cycles. RDF is a valuable tool for examining structural changes within a material [63]. In the 2nd cycle, the probability of finding another particle at a particular distance differs for each cooling rate. However, by the 10th cycle, the probability of finding another particle remains nearly unchanged, suggesting that the interatomic distances between the atoms remain consistent. This indicates that the structural changes occurring within the system are relatively similar as the number of cycles increases.

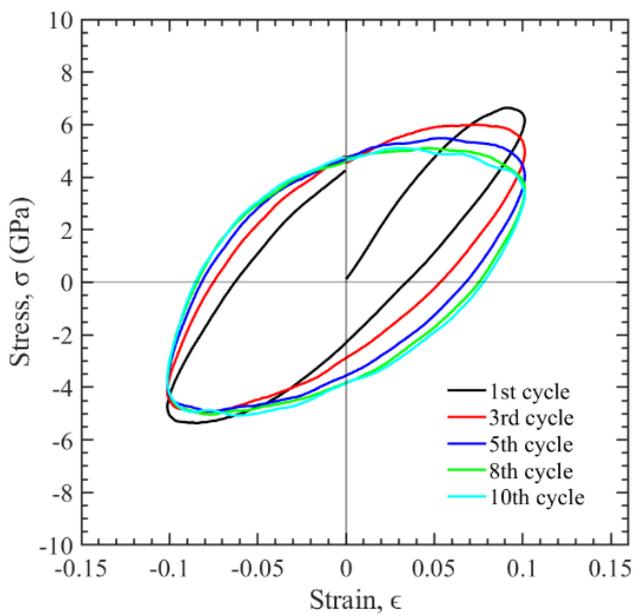

(a)

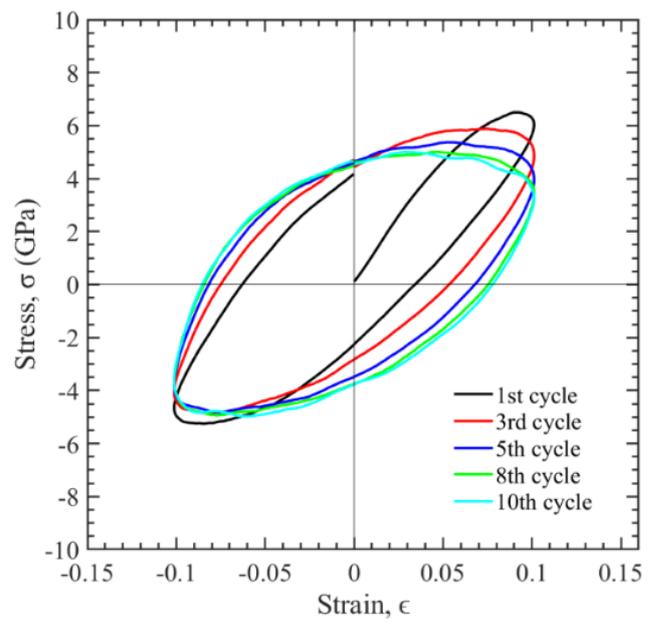

(b)

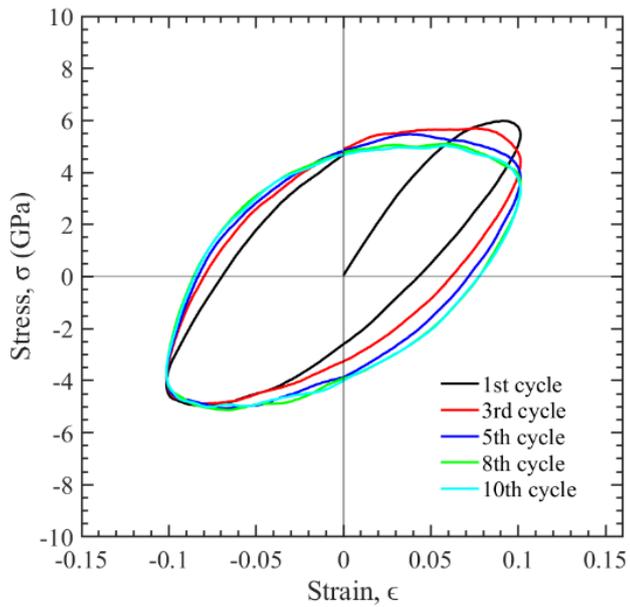

(c)

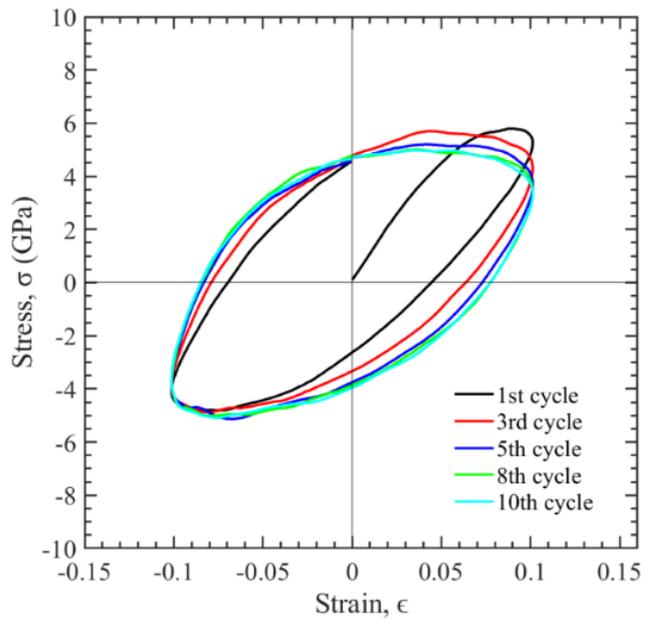

(d)

**Figure 11**: Stress vs strain curves for Sn during cyclic loading when cooled at (a) 2.5K/ps, (b) 10 K/ps, (c) 50 K/ps, and (d) 100 K/ps

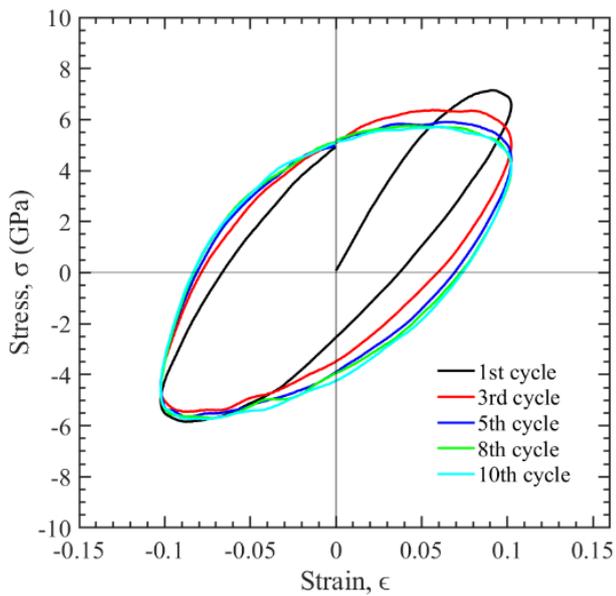

(a)

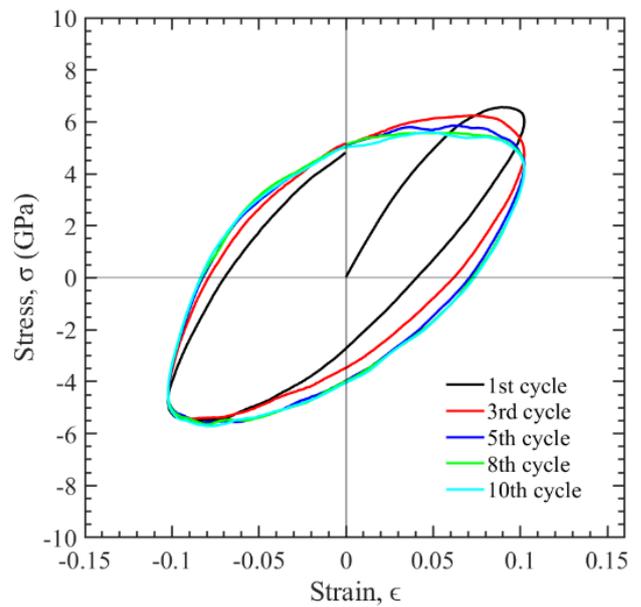

(b)

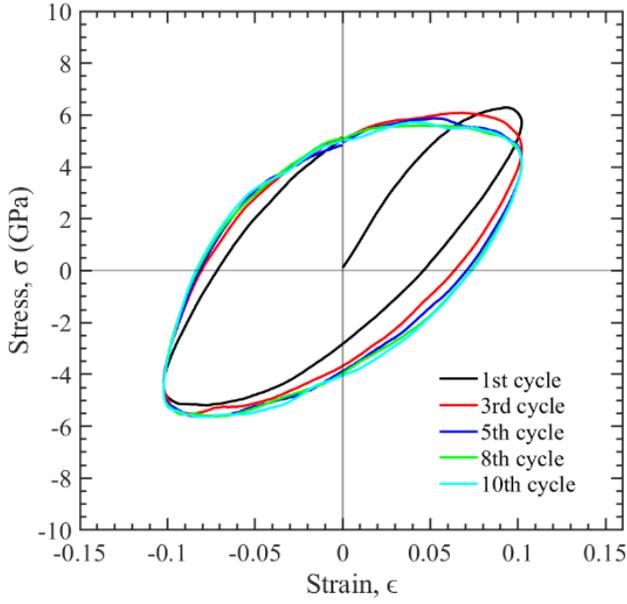

(c)

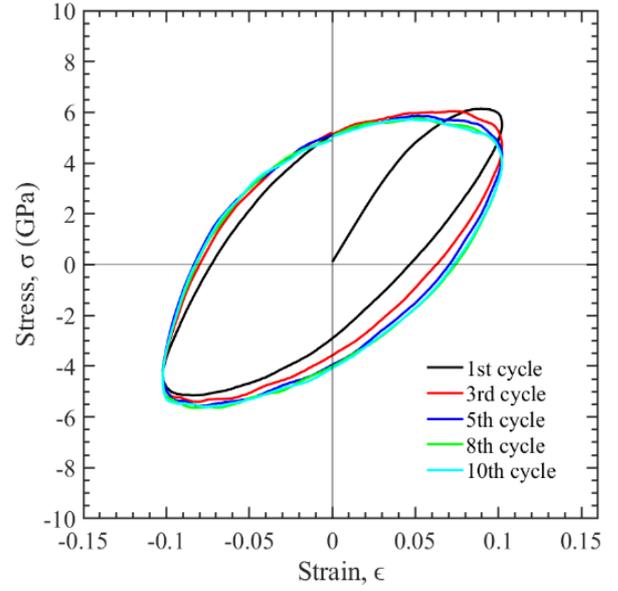

(d)

**Figure 12**: Stress vs strain curves for Sn-Ag during cyclic loading when cooled at (a) 2.5 K/ps, (b) 10 K/ps, (c) 50 K/ps, and (d) 100 K/ps

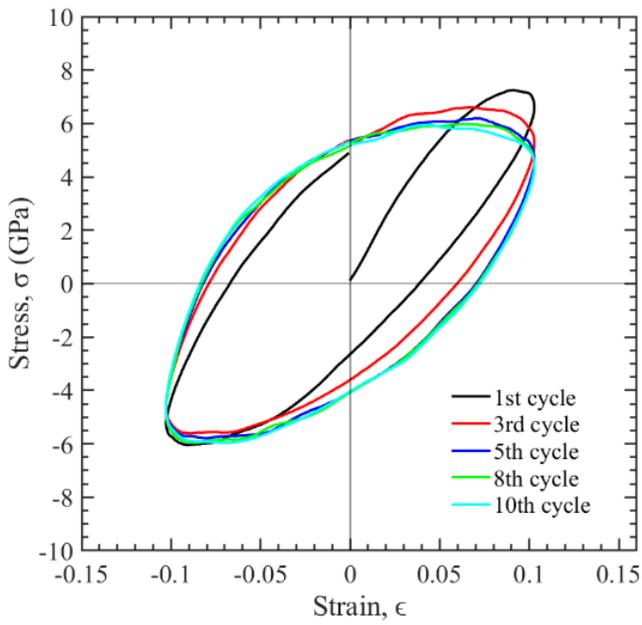

(a)

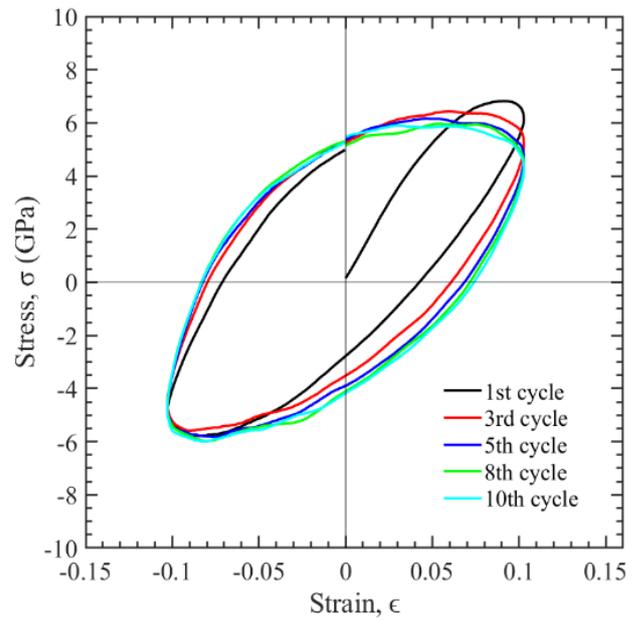

(b)

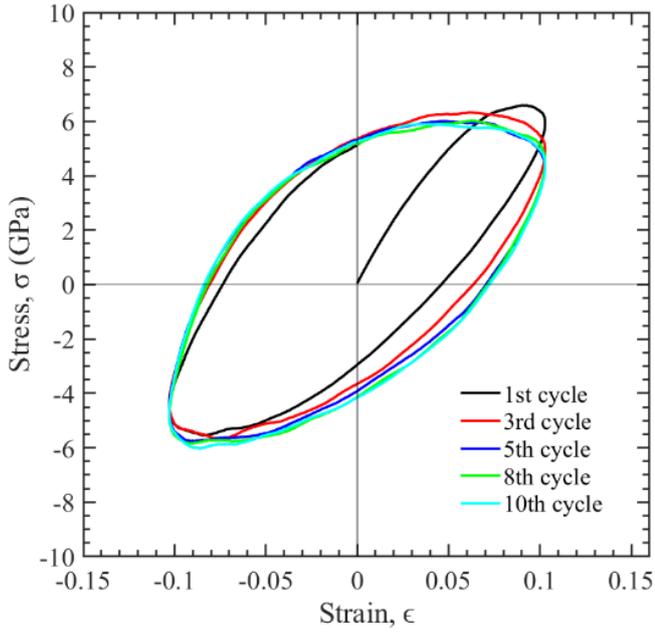

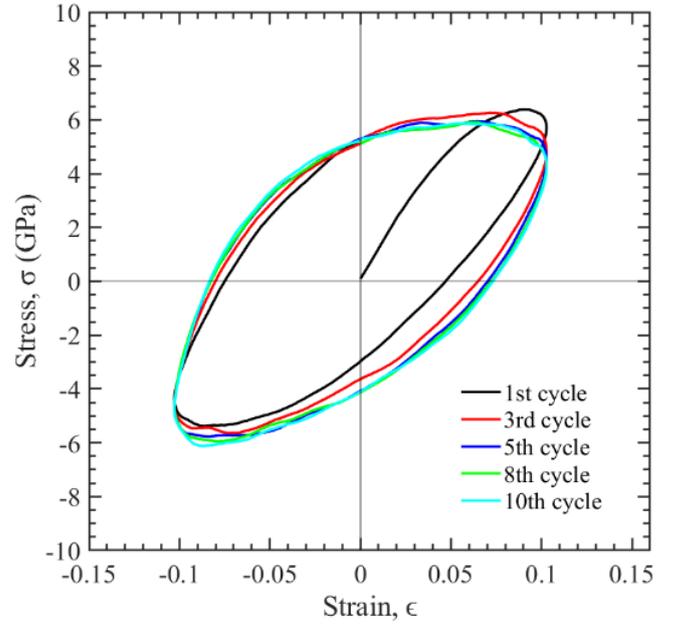

(c)                               (d)

**Figure 13**: Stress vs strain curves for SAC305 during cyclic loading when cooled at (a) 2.5 K/ps, (b) 10 K/ps, (c) 50 K/ps, and (d) 100 K/ps

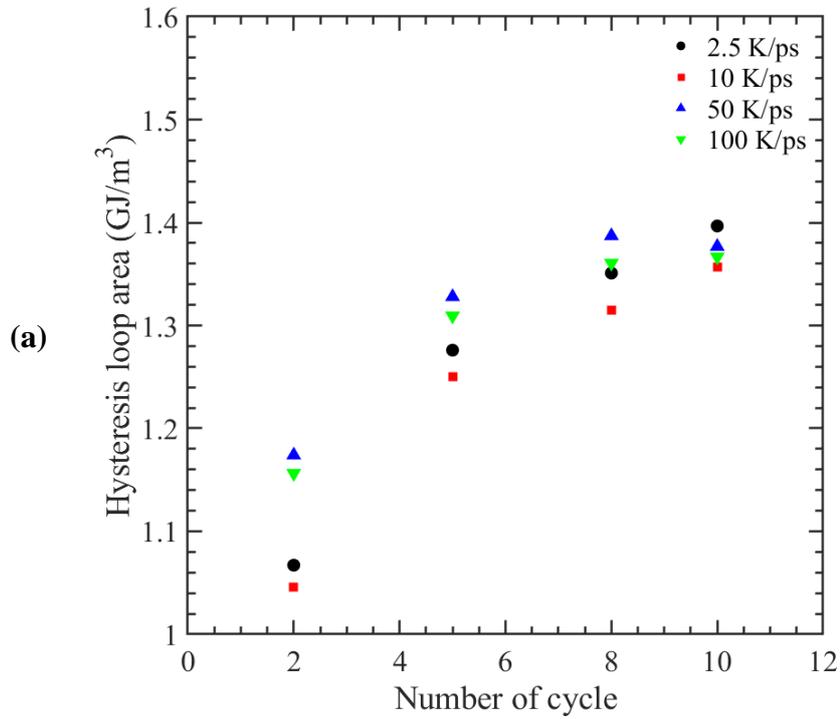

(a)

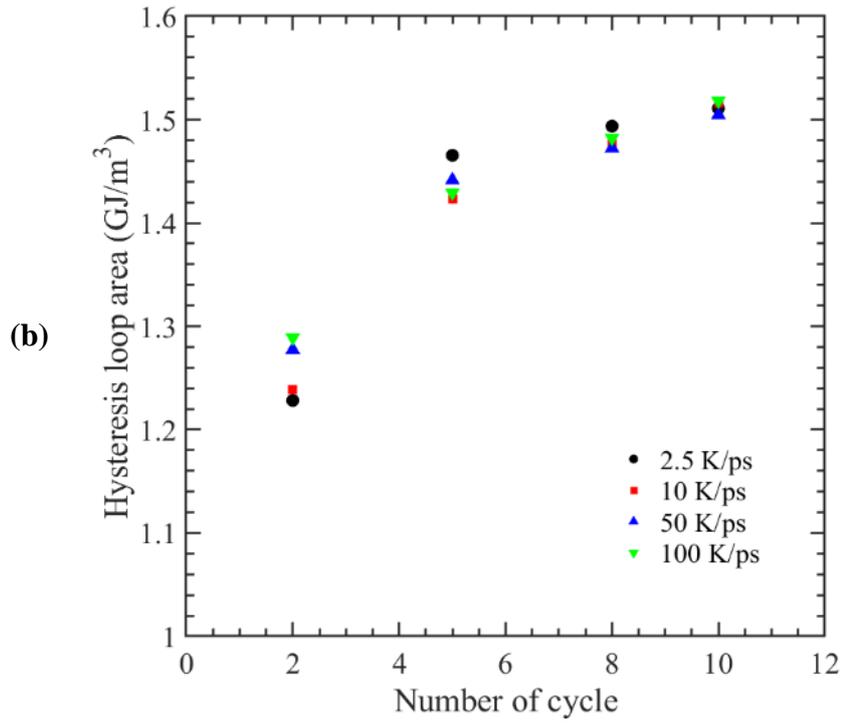

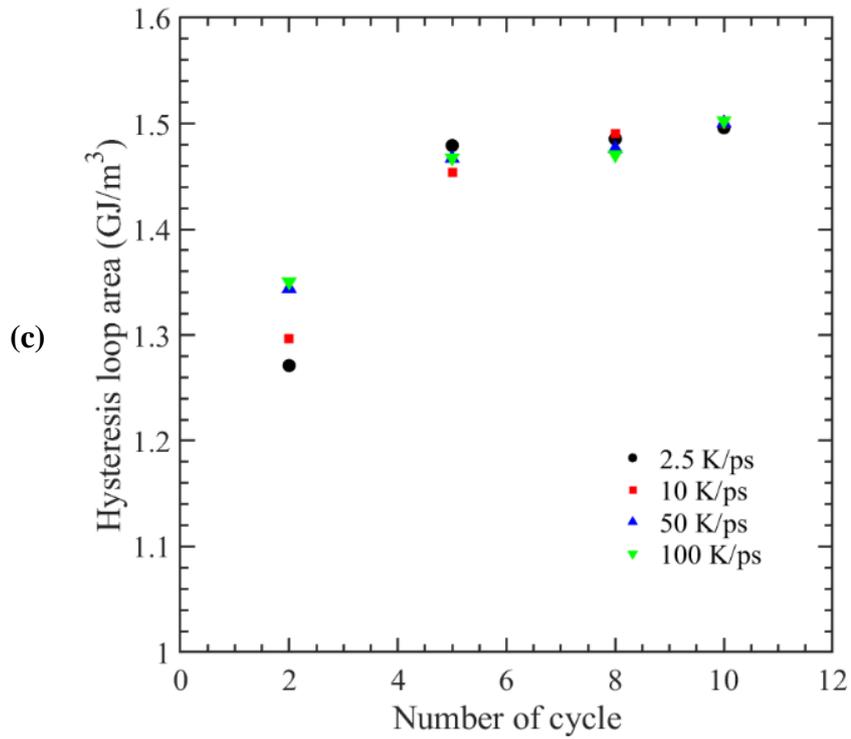

**Figure 14**: Variation of the area of hysteresis loop for different cooling rates of (a) Sn, (b) Sn-Ag, and (c) SAC305

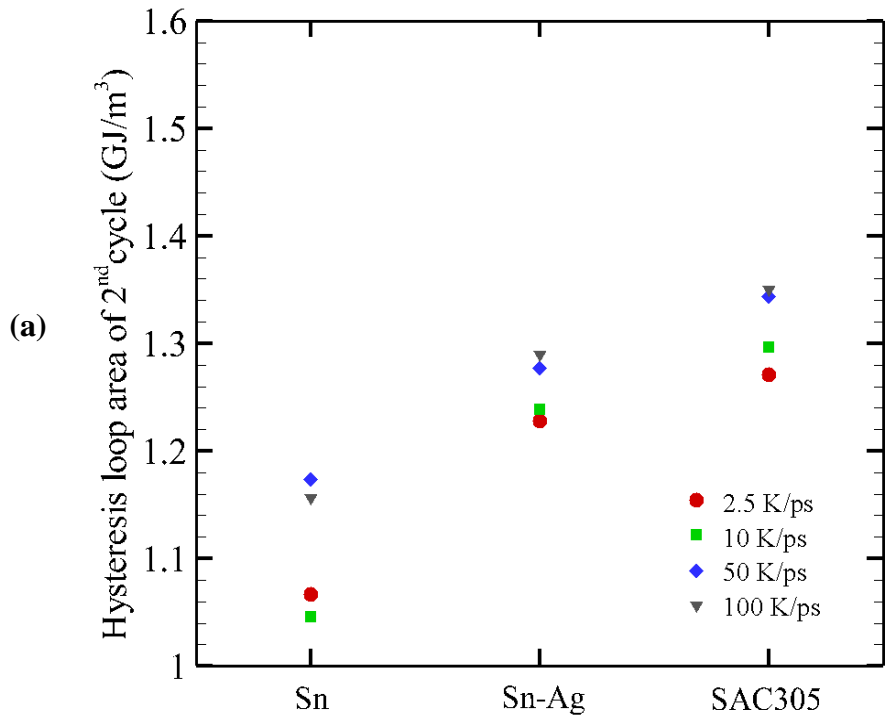

(a)

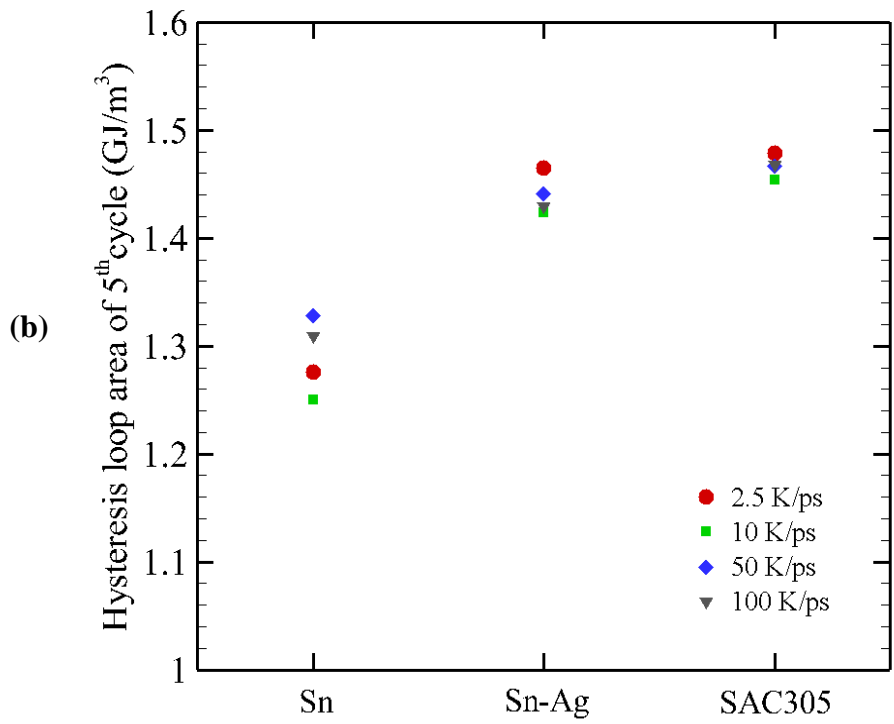

(b)

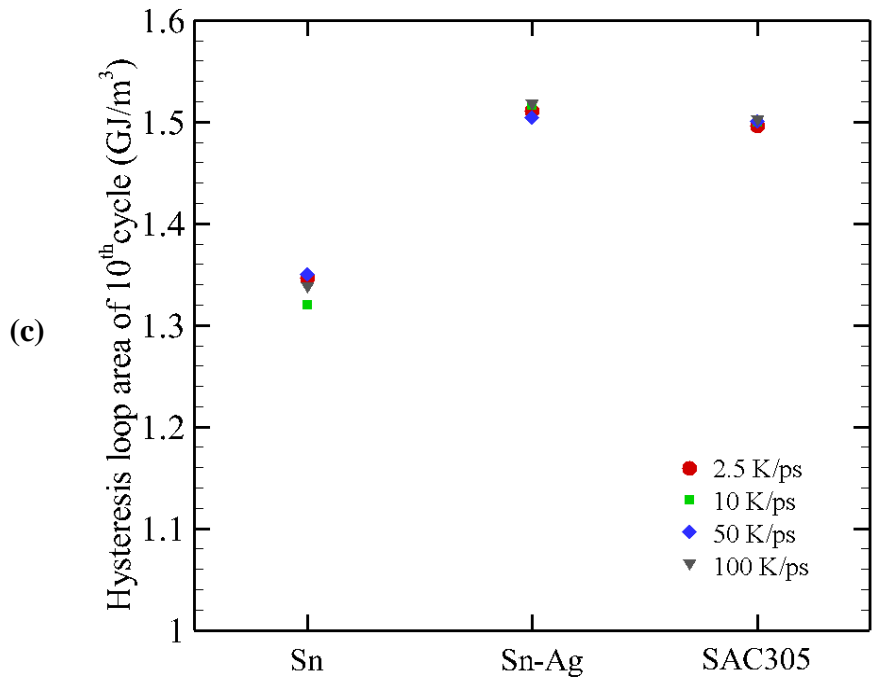

**(c)**

Figure 15: Comparison among the materials in terms of loop area at (a) 2$^{nd}$, (b) 5$^{th}$, and (c) 10$^{th}$ cycle

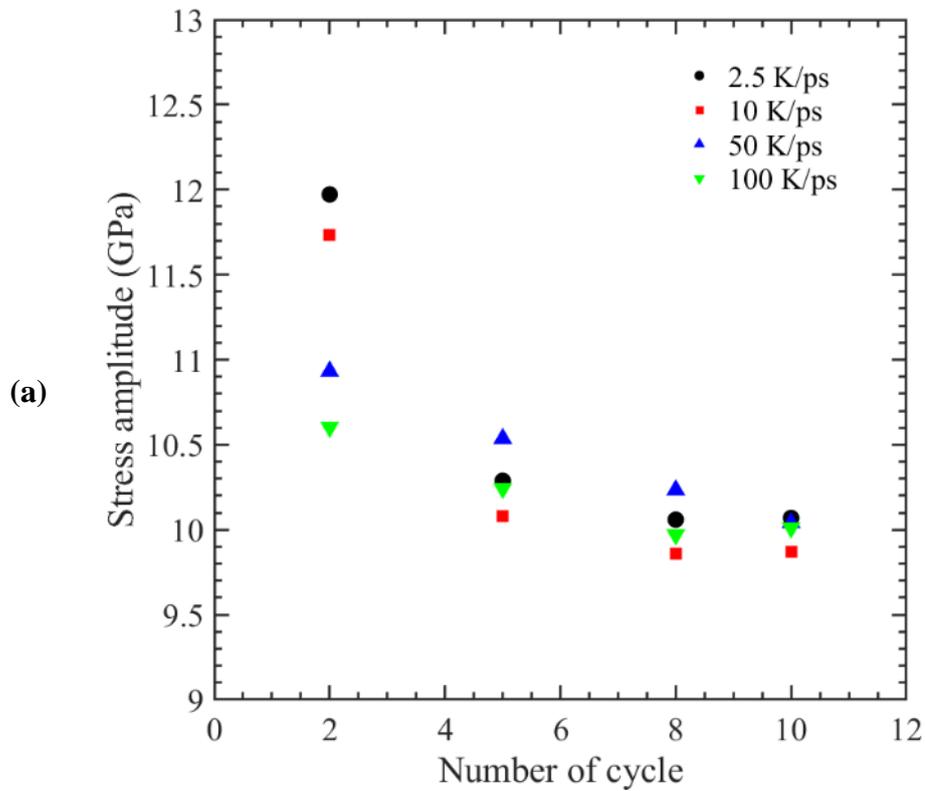

**(a)**

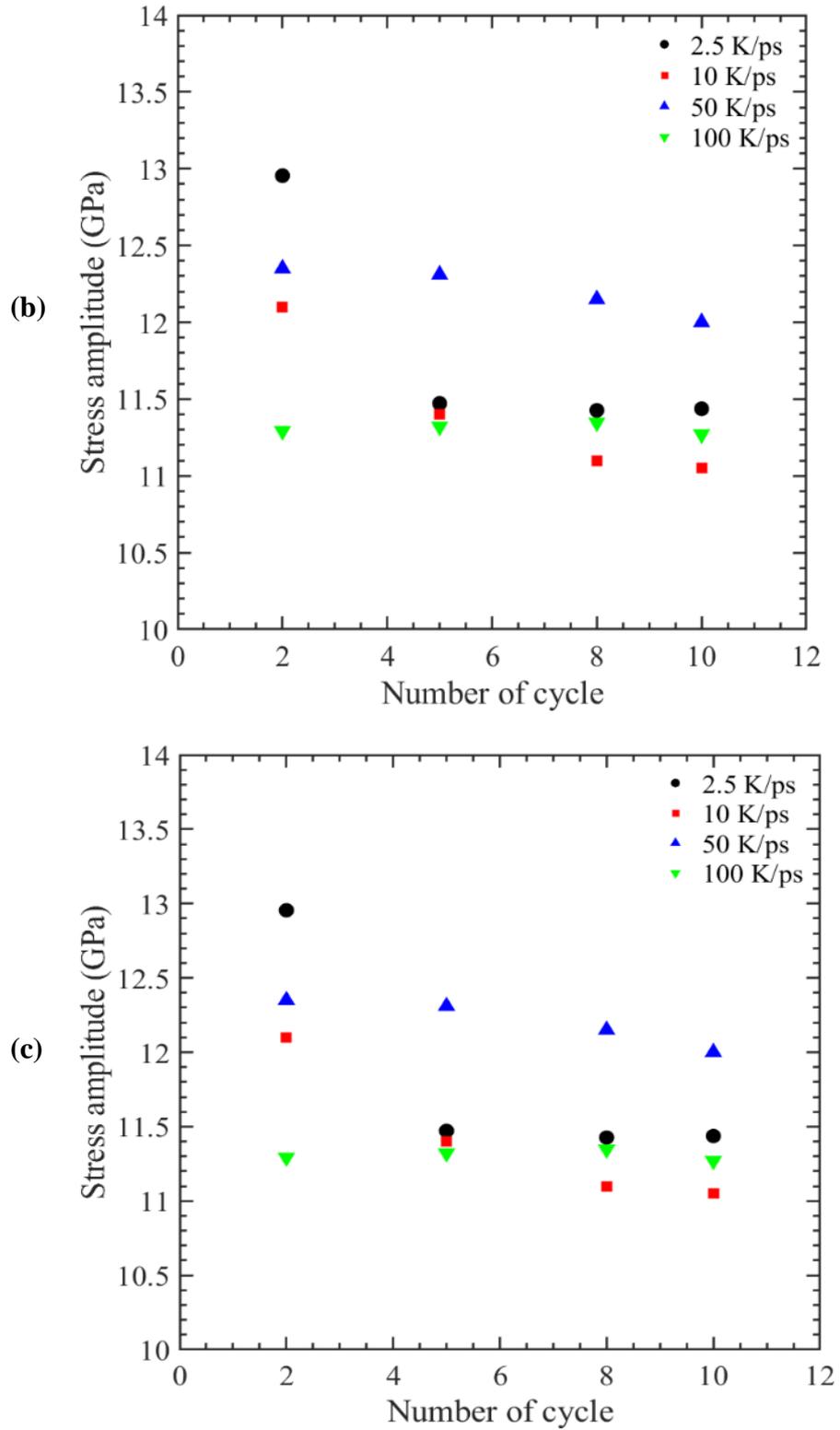

**Figure 16**: Variation of the stress amplitude for different cooling rates for (a) Sn, (b) Sn-Ag, and (c) SAC305

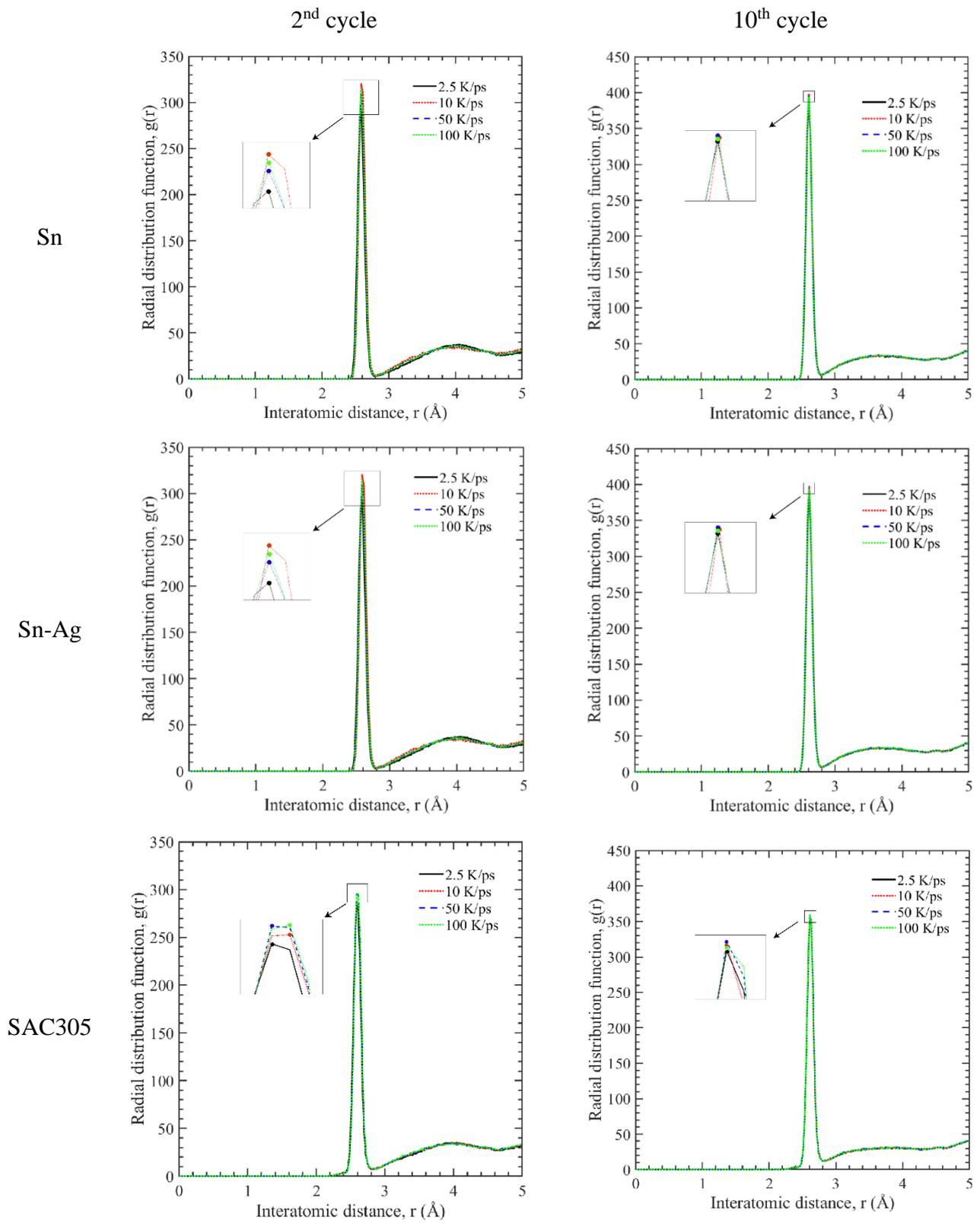

**Figure 17**: Variation of RDF for different cooling rates of Sn, Sn-Ag and SAC305 during 2$^{nd}$ cycle and 10$^{th}$ cycle

## 5. CONCLUSION

In this research, molecular dynamics was employed to model and simulate changes in the mechanical and thermal characteristics of lead-free solder materials at the nanoscale. This investigation offers valuable insights for enhancing solder joint quality and reliability in electronic assemblies, ensuring sound solidification, and mitigating potential defects. Atomistic simulations were conducted to explore the influence of cooling rates on materials subjected to both uniaxial tensile and cyclic loading. The findings indicate the following:

I. An increase in cooling rate leads to a consistent decline in the ultimate tensile strength and Young's modulus across all lead-free solder alloy cases, primarily due to an increase in the number of defects.

II. The modulus of toughness experiences an increasing trend with increasing cooling rates from 2.5 K/ps to 100 K/ps, while the modulus of resilience exhibits a decreasing trend across all materials.

III. Higher cooling rates result in decreased coefficients of thermal expansion in solder alloys, as they have less time to contract.

IV. During cyclic loading, damage accumulation increases with the number of cycles for lead-free solders. Hysteresis loop areas of all lead-free solder materials converge at the 10th cycle, indicating stability in damage accumulation.

V. Initially, during cyclic loading, there are variations in interatomic distances, but as the number of cycles increases, they eventually stabilize and remain consistent.

**ACKNOWLEDGEMENT**

We would like to express our sincere appreciation to the Bangladesh University of Engineering and Technology (BUET) for their exceptional support in providing a high-quality computer laboratory, which was instrumental in the successful completion of this research project.

**REFERENCES**

[1]   J. M. Davis, R. W. Elias, and L. Grant, "Current issues in human lead exposure and regulation of lead," *Neurotoxicology,* vol. 14, no. 2-3, pp. 15-27, 1993.

[2] K. J. Puttlitz and K. A. Stalter, "Handbook of lead-free solder technology for microelectronic assemblies". *CRC Press*, 2004.

[3] O. Salmela, "Acceleration factors for lead-free solder materials," *IEEE transactions on components and packaging technologies,* vol. 30, no. 4, pp. 700-707, 2007.

[4] C. Wu, D. Q. Yu, C. Law, and L. Wang, "Properties of lead-free solder alloys with rare earth element additions," *Materials Science and Engineering: R: Reports,* vol. 44, no. 1, pp. 1-44, 2004.

[5] H. R. Kotadia, P. D. Howes, and S. H. Mannan, "A review: On the development of low melting temperature Pb-free solders," *Microelectronics Reliability,* vol. 54, no. 6-7, pp. 1253-1273, 2014.

[6] E. E. Mhd Noor, N. F. Mhd Nasir, and S. R. A. Idris, "A review: lead free solder and its wettability properties," *Soldering & Surface Mount Technology,* vol. 28, no. 3, pp. 125-132, 2016.

[7] M. McCormack, S. Jin, G. Kammlott, and H. Chen, "New Pb-free solder alloy with superior mechanical properties," *Applied Physics Letters,* vol. 63, no. 1, pp. 15-17, 1993.

[8] A. Schubert, H. Walter, R. Dudek, B. Michel, G. Lefranc, J. Otto, and G. Mitic, "Thermo-mechanical properties and creep deformation of lead-containing and lead-free solders," *Proceedings International Symposium on Advanced Packaging Materials Processes, Properties and Interfaces*, pp. 129-134, 2001.

[9] K. Suganuma, "Lead-free soldering in electronics: Science, technology, and environmental impact". *CRC Press*, 2003.

[10] J. Glazer, "Microstructure and mechanical properties of Pb-free solder alloys for low-cost electronic assembly: a review," *Journal of Electronic Materials,* vol. 23, pp. 693-700, 1994.

[11] Y. Li, K.-s. Moon, and C. Wong, "Electronics without lead," *Science,* vol. 308, no. 5727, pp. 1419-1420, 2005.

[12] M. Abtew and G. Selvaduray, "Lead-free Solders in Microelectronics," *Materials Science and Engineering: R: Reports,* vol. 27, no. 5, pp. 95-141, 2000.

[13] S. Cheng, C.-M. Huang, and M. Pecht, "A review of lead-free solders for electronics applications," *Microelectronics Reliability,* vol. 75, pp. 77-95, 2017.

[14] S. K. Kang, P. Lauro, D. Y. Shih, D. W. Henderson, and K. J. Puttlitz, "Microstructure and mechanical properties of lead-free solders and solder joints used in microelectronic

applications," *IBM Journal of Research and Development,* vol. 49, no. 4.5, pp. 607-620, 2005.

[15] M. Abtew and G. Selvaduray, "Lead-free solders in microelectronics," *Materials Science and Engineering: R: Reports,* vol. 27, no. 5-6, pp. 95-141, 2000.

[16] A. Kroupa *et al.*, "Current problems and possible solutions in high-temperature lead-free soldering," *Journal of Materials Engineering and Performance,* vol. 21, pp. 629-637, 2012.

[17] D. Suraski and K. Seelig, "The current status of lead-free solder alloys," *IEEE Transactions on Electronics Packaging Manufacturing,* vol. 24, no. 4, pp. 244-248, 2001.

[18] O. R. Adetunji, R. A. Ashimolowo, P. O. Aiyedun, O. M. Adesusi, H. O. Adeyemi, and O. R. Oloyede, "Tensile, hardness and microstructural properties of Sn-Pb solder alloys," *Materials Today: Proceedings,* vol. 44, pp. 321-325, 2021.

[19] I. Shohji, T. Yoshida, T. Takahashi, and S. Hioki, "Tensile properties of Sn–Ag based lead-free solders and strain rate sensitivity," *Materials Science and Engineering: A,* vol. 366, no. 1, pp. 50-55, 2004.

[20] M.-L. Wu and J.-S. Lan, "Reliability and failure analysis of SAC 105 and SAC 1205N lead-free solder alloys during drop test events," *Microelectronics Reliability,* vol. 80, pp. 213-222, 2018.

[21] H. L. J. Pang, Y. P. Wang, X. Q. Shi, and Z. P. Wang, "Sensitivity study of temperature and strain rate dependent properties on solder joint fatigue life," in *Proceedings of 2nd Electronics Packaging Technology Conference (Cat. No.98EX235)*, 1998, pp. 184-189.

[22] P. Lall, V. Yadav, J. Suhling, and D. Locker, "Low Temperature Material Characterization of Lead-Free SAC Solder Alloy at High Strain Rate After Prolonged High Temperature Storage," in *ASME 2021 International Technical Conference and Exhibition on Packaging and Integration of Electronic and Photonic Microsystems*, pp. V001T07A021, 2021.

[23] M. R. Chowdhury, S. Ahmed, A. Fahim, J. C. Suhling and P. Lall, "Mechanical characterization of doped SAC solder materials at high temperature," *2016 15th IEEE Intersociety Conference on Thermal and Thermomechanical Phenomena in Electronic Systems (ITherm)*, Las Vegas, NV, USA, pp. 1202-1208, 2016.

[24] C. Basaran, A. Cartwright, and Y. Zhao, "Experimental Damage Mechanics of Microelectronics Solder Joints under Concurrent Vibration and Thermal Loading," *International Journal of Damage Mechanics,* vol. 10, no. 2, pp. 153-170, 2001.

[25] S. Wiese and S. Rzepka, "Time-independent elastic–plastic behaviour of solder materials," *Microelectronics Reliability,* vol. 44, no. 12, pp. 1893-1900, 2004.

[26] A. Schubertt, R. Dudek, E. Auerswald, A. Gollbardt, B. Michel, and H. Reichl, "Fatigue life models for SnAgCu and SnPb solder joints evaluated by experiments and simulation," *Electronic components and technology conference,* pp. 603-610, 2003.

[27] M. Motalab, R. Paul, S. Saha, S. Mojumder, M. Ahmed, and J. Suhling, "Atomistic analysis of the thermomechanical properties of Sn–Ag–Cu solder materials at the nanoscale with the MEAM potential," *Journal of Molecular Modeling,* vol. 25, pp. 1-10, 2019.

[28] L. Zhang, D. Xiong, J. Li, L. Yin, Z. Yao, G. Wang, L. Zhang, and H. Zhang., "Molecular dynamics simulation of the interfacial evolution and whisker growth of copper-tin coating under electrothermal coupling," *Computational Materials Science,* vol. 202, p. 110981, 2022.

[29] J. Zhang, J. Yang, L. Liang, Y. Xu, and J. Guo, "A molecular dynamics investigation of the micro-mechanism for vacancy formation between $Ag_3Sn$ and βSn under electromigration," *Molecular Physics,* vol. 116, no. 1, pp. 99-106, 2018.

[30] Y. Li, L. Xu, L. Zhao, K. Hao, and Y. Han, "Shear deformation behavior and failure mechanisms of graphene reinforced Sn-based solder joints bonded by transient current," *Materials & Design,* vol. 224, p. 111369, 2022.

[31] B. Chen, W.-P. Wu, M.-X. Chen, and Y.-F. Guo, "Molecular dynamics study of fatigue mechanical properties and microstructural evolution of Ni-based single crystal superalloys under cyclic loading," *Computational Materials Science,* vol. 185, p. 109954, 2020.

[32] S. M. A. A. Alvi, A. Faiyad, M. A. M. Munshi, M. Motalab, M. M. Islam, and S. Saha, "Cyclic and tensile deformations of Gold–Silver core shell systems using newly parameterized MEAM potential," *Mechanics of Materials,* vol. 169, p. 104304, 2022.

[33] H. G. Nguyen, T. H. Fang, and D. Q. Doan, "Cyclic plasticity and deformation mechanism of AlCrCuFeNi high entropy alloy," *Journal of Alloys and Compounds,* vol. 940, p. 168838, 2023/04/15/ 2023.

[34] G. S. Grest, D. J. Srolovitz, and M. P. Anderson, "Computer simulation of grain growth—IV. Anisotropic grain boundary energies," *Acta Metallurgica,* vol. 33, no. 3, pp. 509-520, 1985.

[35]   R. W. Lewis and P. M. Roberts, "Finite element simulation of solidification problems," in *Modelling the Flow and Solidification of Metals*, T. J. Smith Ed. Dordrecht: Springer Netherlands, pp. 61-92, 1987.

[36]   E. Miyoshi and T. Takaki, "Extended higher-order multi-phase-field model for three-dimensional anisotropic-grain-growth simulations," *Computational Materials Science,* vol. 120, pp. 77-83, 2016.

[37]   A. J. Haslam, S. R. Phillpot, D. Wolf, D. Moldovan, and H. Gleiter, "Mechanisms of grain growth in nanocrystalline fcc metals by molecular-dynamics simulation," *Materials Science and Engineering: A,* vol. 318, no. 1, pp. 293-312, 2001.

[38]   Y. Shibuta, S. Sakane, E. Miyoshi, S. Okita, T. Takaki, and M. Ohno, "Heterogeneity in homogeneous nucleation from billion-atom molecular dynamics simulation of solidification of pure metal," *Nature Communications,* vol. 8, no. 1, p. 10, 2017.

[39]   J. W. Elmer, S. M. Allen, and T. W. Eagar, "Microstructural development during solidification of stainless steel alloys," *Metallurgical Transactions A,* vol. 20, no. 10, pp. 2117-2131, 1989.

[40]   A. Mahata, M. A. Zaeem, and M. I. Baskes, "Understanding homogeneous nucleation in solidification of aluminum by molecular dynamics simulations," *Modelling and Simulation in Materials Science and Engineering,* vol. 26, no. 2, p. 025007, 2018.

[41]   J. Li *et al.*, "Tuning the mechanical behavior of high-entropy alloys via controlling cooling rates," *Materials Science and Engineering: A,* vol. 760, pp. 359-365, 2019.

[42]   B. Shen, C.Y. Liu, Y. Jia, G.Q. Yue, F.S. Ke, H.B. Zhao, L.Y. Chen, S.Y. Wang, C.Z. Wang, and K.M. Ho, "Molecular dynamics simulation studies of structural and dynamical properties of rapidly quenched Al," *Journal of Non-Crystalline Solids,* vol. 383, pp. 13-20, 2014.

[43]   S. Li, S. Cui, H. Chen, J. Li, H. Huang, and H. Luo, "Effect of cooling rates on solidification, microstructure and mechanical properties in tungsten," *CrystEngComm,* vol. 21, no. 26, pp. 3930-3938, 2019.

[44]   T. Shu, N. Hu, F. Liu, and G. J. Cheng, "Nanoparticles induced intragranular and dislocation substructures in powder bed fusion for strengthening of high-entropy-alloy," *Materials Science and Engineering: A,* vol. 878, p. 145110, 2023.

[45]   M. I. Baskes, "Modified embedded-atom potentials for cubic materials and impurities," *Physical Review B,* vol. 46, no. 5, pp. 2727-2742, 1992.

[46] M. I. Baskes, J. S. Nelson, and A. F. Wright, "Semiempirical modified embedded-atom potentials for silicon and germanium," *Physical Review B,* vol. 40, no. 9, pp. 6085-6100, 1989.

[47] H. Fei, K. Yazzie, J. Williams, and H. Jiang, "Multiscale Modeling of the Interfacial Fracture Behavior in the Sn-Cu6 Sn5-Cu System," *Journal of Computational and Theoretical Nanoscience,* vol. 8, pp. 873-880, 2011.

[48] H. Dong, L. Fan, K.S. Moon, C. P. Wong, and M. Baskes, "MEAM molecular dynamics study of lead free solder for electronic packaging applications," *Modelling and Simulation in Materials Science and Engineering,* vol. 13, p. 1279, 2005.

[49] R. Ravelo and M. Baskes, "Equilibrium and Thermodynamic Properties of Grey, White, and Liquid Tin," *Physical Review Letters,* vol. 79, no. 13, pp. 2482-2485, 1997.

[50] S. Plimpton, "Fast Parallel Algorithms for Short-Range Molecular Dynamics," *Journal of Computational Physics,* vol. 117, no. 1, pp. 1-19, 1995.

[51] A. Stukowski, "Visualization and analysis of atomistic simulation data with OVITO–the Open Visualization Tool," *Modelling and Simulation in Materials Science and Engineering,* vol. 18, no. 1, p. 015012, 2010.

[52] M. Mustafa, Z. Cai, J. C. Suhling, and P. Lall, "The effects of aging on the cyclic stress-strain behavior and hysteresis loop evolution of lead free solders," in *2011 IEEE 61st Electronic Components and Technology Conference (ECTC)*, pp. 927-939, 2011.

[53] M. Zhou, "A new look at the atomic level virial stress: on continuum-molecular system equivalence," *Proceedings of the Royal Society of London. Series A: Mathematical, Physical and Engineering Sciences,* vol. 459, no. 2037, pp. 2347-2392, 2003.

[54] E. George, D. Das, M. Osterman, and M. Pecht, "Thermal Cycling Reliability of Lead-Free Solders (SAC305 and Sn3.5Ag) for High-Temperature Applications," *IEEE Transactions on Device and Materials Reliability,* vol. 11, no. 2, pp. 328-338, 2011.

[55] E. V. Vernon and S. Weintroub, "The Measurement of the Thermal Expansion of Single Crystals of Indium and Tin with a Photoelectric Recording Dilatometer," *Proceedings of the Physical Society. Section B,* vol. 66, no. 10, p. 887, 1953.

[56] A. Schubert, H. Walter, R. Dudek, B. Michel, G. Lefranc, J. Otto, and G Mitic , "Thermo-mechanical properties and creep deformation of lead-containing and lead-free solders," *Proceedings International Symposium on Advanced Packaging Materials Processes, Properties and Interfaces (IEEE Cat. No.01TH8562)*, pp. 129-134, 2001.

[57] F. X. Che, E. C. Poh, W. H. Zhu, and B. S. Xiong, "Ag Content Effect on Mechanical Properties of Sn-xAg-0.5Cu Solders," in *2007 9th Electronics Packaging Technology Conference*, pp. 713-718, 2007.

[58] K. Newman, "BGA brittle fracture - alternative solder joint integrity test methods," *Proceedings Electronic Components and Technology,* pp. 1194-1201 Vol. 2, 2005.

[59] F. Song, S. W. R. Lee, K. Newman, B. Sykes, and S. Clark, "Brittle Failure Mechanism of SnAgCu and SnPb Solder Balls during High Speed Ball Shear and Cold Ball Pull Tests," *Proceedings 57th Electronic Components and Technology Conference*, 29, pp. 364-372, 2007.

[60] M. Mustafa, Z. Cai, J. Suhling, and P. Lall, "The effects of aging on the cyclic stress-strain behavior and hysteresis loop evolution of lead free solders," *Proceedings - Electronic Components and Technology Conference,* pp. 927-939, 2011.

[61] T. L. Jesse and L. M. David, "Degradation of an Ni-Ti alloy during cyclic loading," in *Proc.SPIE*, vol. 2189, pp. 326-341, 1994.

[62] G. Kang and Y. Liu, "Uniaxial ratchetting and low-cycle fatigue failure of the steel with cyclic stabilizing or softening feature," *Materials Science and Engineering: A,* vol. 472, no. 1, pp. 258-268, 2008.

[63] D. Srolovitz, T. Egami, and V. Vitek, "Radial distribution function and structural relaxation in amorphous solids," *Physical Review B,* vol. 24, no. 12, pp. 6936-6944, 1981.